\documentclass{aa}  

\usepackage{graphicx}
\usepackage{txfonts}

\newcommand{\quotes}[1]{``#1''}
\newcommand\code[1]{\textsc{\MakeLowercase{#1}}}

\def\be{\begin{equation}}
\def\ee{\end{equation}}

\def\kms{km s$^{-1}$}

\def\msun{{\rm M}_{\odot}}
\def\zsun{{\rm Z}_{\odot}}

\def\msunyr{\msun/{\rm yr}}
\def\msunpc2{\msun/{\rm pc}^{2}}

\def\gsim{\lower.5ex\hbox{\gtsima}} 
\def\lsim{\lower.5ex\hbox{\ltsima}} 
\def\gtsima{$\; \buildrel > \over \sim \;$} 
\def\ltsima{$\; \buildrel < \over \sim \;$} \def\gsim{\lower.5ex\hbox{\gtsima}} 
\def\lsim{\lower.5ex\hbox{\ltsima}} 
\def\simgt{\lower.5ex\hbox{\gtsima}} 
\def\simlt{\lower.5ex\hbox{\ltsima}} 

\def\CII{\hbox{[C~$\scriptstyle\rm II $]}}
\def\OIII{\hbox{[O~$\scriptstyle\rm III $]}}

\def\kms{{\rm km\, s^{-1}}}

\def\dust{\mathcal{D}}
\def\dsun{\dust_{\odot}}

\def\sigmacii{\sigma_{\rm[CII]}}
\def\sigmahalpha{\sigma_{\rm H\alpha}}
\def\sigmaline{\sigma_{\rm em-line}}
\def\sigmant{\sigma_{\rm nt}}
\def\sigmath{\sigma_{\rm th}}

\usepackage[T1]{fontenc}

\usepackage{aecompl}
\usepackage{graphicx}	
\usepackage{amsmath}	
\usepackage{amssymb}	
\usepackage{newtxtext,newtxmath}

\usepackage{subfig}
\usepackage{comment}
\usepackage{multirow}

\usepackage{mathrsfs}

\usepackage{multirow}
\usepackage{graphicx}
\usepackage{txfonts}
\usepackage[dvipsnames]{xcolor}
\usepackage{hyperref}
\usepackage{booktabs}
\usepackage{verbatim} 
 
\hypersetup{
    colorlinks=true,
    linkcolor=blue,
    filecolor=magenta,      
    urlcolor=blue,
    pdftitle={Overleaf Example},
    pdfpagemode=FullScreen,
    citecolor=blue
    }
\usepackage{orcidlink}
\begin{document} 

\title{Dynamically cold disks in the early Universe: myth or reality?}
\titlerunning{Dynamically cold disks in the early Universe}
\author{
        M. Kohandel\inst{1}\thanks{\email{mahsa.kohandel@sns.it}}\orcidlink{0000-0003-1041-7865},
        A. Pallottini\inst{1}\orcidlink{0000-0002-7129-5761},
        A. Ferrara \inst{1}\orcidlink{0000-0002-9400-7312},
        A. Zanella\inst{2}\orcidlink{0000-0001-8600-7008},
        F. Rizzo \inst{3,4}\orcidlink{0000-0001-9705-2461},
        S. Carniani\inst{1}\orcidlink{0000-0002-6719-380X}
       }

\institute{
           Scuola Normale Superiore, Piazza dei Cavalieri 7, I-56126 Pisa, Italy \and
           INAF - Osservatorio Astronomico di Padova, Vicolo Osservatorio 5, 35122, Padova, Italy \and
           Cosmic Dawn Center (DAWN) \and
           Niels Bohr Institute, University of Copenhagen, Jagtvej 128, 2200 Copenhagen, Denmark
           }
\authorrunning{M. Kohandel et al.}
\date{Received October XX, 2023; accepted XXX XX, XXXX}

\abstract
  {Theoretical models struggle to reproduce dynamically cold disks with significant rotation-to-dispersion support ($V_{\rm{rot}}/\sigma$) observed in star-forming galaxies in the early Universe, at redshift $z>4$.}
  {We aim to explore the possible emergence of dynamically cold disks in cosmological simulations and to understand if different kinematic tracers can help reconcile the tension between theory and observations.}
  {We use 3218 galaxies from the \code{serra} suite of zoom-in simulations, with   $8\le\log{(M_\star/M_\odot)}\le 10.3$ and $\rm{SFR} \le 128\,\msunyr$, within $4\le z \le 9$ range. We generate hyper-spectral data cubes for $2\times3218$ synthetic observations of H$\alpha$ and [CII].
  }
  {We find that the choice of kinematic tracer strongly influences gas velocity dispersion ($\sigma$) estimates. When using H$\alpha$ (\CII) synthetic observations, we observe a strong (mild) correlation between $\sigma$ and $M_\star$. Such a difference arises mostly for $M_\star > 10^{9}\ \msun$ galaxies, for which $\sigmahalpha > 2 \sigmacii$ for a significant fraction of the sample. Regardless of the tracer, our predictions suggest the existence of massive ($M_\star > 10^{10} \ \msun$) galaxies with $V_{\rm{rot}}/\sigma > 10$ at $z>4$, maintaining cold disks for $>10$ orbital periods ($\sim 200\,\rm{Myr}$). Furthermore, we do not find any significant redshift dependence for $V_{\rm{rot}}/\sigma$ ratio in our sample. 
  }
  {Our simulations predict the existence of dynamically cold disks in the early Universe. However, different tracers are sensitive to different kinematic properties. While \CII~effectively traces the thin, gaseous disk of galaxies, H$\alpha$ includes the contribution from ionized gas beyond the disk region, characterized by prevalent vertical or radial motions that may be associated with outflows. We show that the presence of H$\alpha$ halos could be a signature of such galactic outflows. This result emphasizes the importance of combining ALMA and JWST/NIRspec studies of high-$z$ galaxies.}


\maketitle

\section{Introduction}

Disks are almost ubiquitous within the star-forming galaxies (SFGs) population in the local Universe. Somewhat surprisingly, recent observations \citep{Ferreira+22a,Kartaltepe+23,Robertson+23,Tohill+23} using the James Webb Space Telescope (JWST) have unveiled their presence also in the very early Universe, stretching as far back as $z\sim 9$. However, these early disks might exhibit distinct dynamical characteristics when compared to their local counterparts.

Indeed, extensive kpc-scale near-infrared Integral-Field-Unit (IFU) observations of star forming galaxies (SFGs) around cosmic noon ($1 \le z \le 3$, e.g., \citealt{Law+09, Stott+16, Forster-schreiber+18, Mieda+16, Mason+17,Turner+17, Wisnioski+19, Birkin+23}) have found a significant increase of the gas velocity dispersion ($\sim 50-100\, \kms$) compared to local SFG values ($\sim 20-25\, \kms$, \citealt{Anderson+06, Epinat+10}).
Notably, the rotation-to-dispersion ratio ($V_{\mathrm{rot}}/\sigma$) for cosmic noon disks typically falls in the range $1-10$ \citep{Law+09, Gnerucci+11, Genzel+11, Johnson+18, Birkin+23}, in contrast to the value of $10-20$ observed in Milky Way and other local spiral disks \citep{Epinat+10}.
As a result, a significant conclusion drawn by these studies is that SFGs become dynamically hotter, featuring substantial pressure support, towards high-$z$.

In recent years, significant progress has been made in characterizing the dynamics of normal SFGs during Cosmic High Noon ($3 \le z \le 6$) and even into the Epoch of Reionization (EoR: $z>6$) through far-infrared (FIR) emission line observations, particularly \CII-158$\mu$m with the Atacama Large Millimeter/Submillimetre Array (ALMA) \citep{Jones+17, Smit+18, Bakx+20, Hashimoto+19, Harikane+20, Herrera-Camus+22, Lefevre+20, Romano+21, Fujimoto+21, Tokuoka+22, Parlanti+23, Posses+23}.
The estimated value of $V_{\mathrm{rot}}/\sigma$ for these galaxies falls in the $\sim 1-7.5$ range. However, these observations are only marginally resolved, leading to a potential underestimate of $V_{\mathrm{rot}}/\sigma$ caused by beam-smearing effects \citep{Kohandel+20,Rizzo+22}.

Recent breakthrough observations \citep{Rizzo+20, Lelli+21, Rizzo+21, Roman-Oliveira+23} achieved higher spatial resolution and sensitivity, robustly characterizing the dynamics of early galaxies. Massive, dusty starburst galaxies ($M_\star > 10^{10}M_{\odot}$) at $z> 4$ exhibit disk structures with $V_{\mathrm{rot}}/\sigma \sim 10$. Moreover, \citealt{Pope+23} have recently revealed stable, rotation-dominated disks ($V_{\mathrm{rot}}/\sigma=5.3\pm 3.6$) in relatively low-mass galaxies ($M_\star \sim 10^{9}\,M_\odot$) during Cosmic High Noon. Adding to this remarkable progress, \citealt{Rowland+23} have recently discovered the most distant ($z\sim 7$) massive ($M_\star > 10^{10}M_{\odot}$) dynamically cold disk ($V_{\mathrm{rot}}/\sigma \sim 10$) in REBELS-25, part of an ALMA large program \citep{Buowens+21}.  

Lastly, JWST/NIRspec multi-object spectroscopy of H$\alpha$ and \OIII~emission lines has started shedding light on the kinematics of ionized gas in EoR galaxies \citep{DeGraaff+23, Parlanti+23b}, hinting that early galaxies can settle into dynamically cold disks (DCD).
Despite this progress, the statistical relevance of DCDs in early SFGs is uncertain, given the limited number of deep spatially resolved observations.  

From a theoretical standpoint, most studies \citep{Dekel+14, Zolotov+15, Hayward+17, Pillepich+19} struggle to explain the existence of DCDs at $z>4$. For instance, \code{tng50} \citep{Pillepich+19} shows an average $V_{\mathrm{rot}}/\sigma < 3$ at $z=4$, similarly to most cosmological simulations.
Nonetheless, a few studies predict that relatively massive galaxies ($M_\star\sim 10^{10}M_\odot$) can temporarily sustain cold disks formed through intense accretion of co-planar, co-rotating gas at $z>3$ \citep{Kretschmer+22}.
Regarding  EoR galaxies, \citealt{Kohandel+20} show that moderate rotation support ($V_{\mathrm{rot}}/\sigma \sim 7$) can be achieved in $M_\star \sim 10^{10}M_\odot$ galaxies as far back as $z\sim 6$, however, this conclusion is based on a relatively small sample of galaxies.

In this {\it Letter}, we exploit the \code{serra} simulations \citep{Pallottini+22} to investigate the kinematic properties of normal SFGs at $4<z<9$ through the analysis of mock observations of  \CII~emission line at $158\,\mu\rm{m}$, a tracer of cold ($T \sim 100\,\rm{K}$) molecular/neutral gas, and H$\alpha$ emission line, a tracer of warm ($T \sim 10^4\,\rm{K}$) ionized gas.

\section{SERRA simulations}\label{sec:serra}

\subsection{Galaxy formation and evolution}

The \code{SERRA} suite of simulations focuses on studying the formation and evolution of galaxies during the EoR \citep[][]{Pallottini+22}. Gas and dark matter are evolved using a customized version of the adaptive mesh refinement code \code{RAMSES} \citep{Teyssier+02}. \code{KROME} \citep{Grassi+14} is employed to model the non-equilibrium chemical network, that includes H, H$^{+}$, H$^{-}$, He, He$^{+}$, He$^{++}$, H${2}$, H${2}^{+}$, electrons, and metals, encompassing $\sim 40$ reactions \citep{bovino:2016aa, pallottini:2017althaea}.
The tracking of metallicity ($Z$) involves summing heavy elements, assuming solar abundance ratios for different metal species \citep{asplund:2009}. Dust is approximated to scale with a fixed dust-to-metal ratio, denoted as $\dust = \dsun (Z/\zsun)$, where $\dsun/\zsun \simeq 0.3$ for the Milky Way \citep[][]{Hirashita+02}. A Milky Way-like grain size distribution is adopted \citep{Weingartner+01}. An initial metallicity floor of $Z_{\rm floor}=10^{-3}\zsun$ is adopted, as expected from pre-enrichment of the intergalactic medium around density peaks \citep[][]{Madau+01, pallottini+14, Pallottini+14b}.

The conversion of molecular hydrogen into stars follows a \citet{schmidt:1959}-\citet{Kennicutt+98}-like relation \citep{pallottini:2017althaea}. These stars, in turn, act as sources of metals, mechanical energy, and radiation \citep{pallottini:2017dahlia}. 
The interstellar radiation field (ISRF) is dynamically evolved on the fly using the moment-based solver from \code{RAMSES-RT} \citep{Rosdahl+13}, which is linked to the chemical evolution \citep{pallottini:2019,decataldo:2019}. To efficiently model radiation propagation, the speed of light is reduced by a factor of $10^3$ in \code{SERRA}, leading to negligible deviations compared to a $10^2$ reduction \citep{pallottini:2019,lupi:2020MNRAS}. Simulations track radiation in 5 energy bins, with one bin partially covering the Habing band ($6.0<{h}\nu <11.2$), one dedicated to the Lyman-Werner band ($11.2<{h}\nu <13.6$) to address H$_2$ photoevaporation, and the remaining three bins covering ionization processes from H to the first ionization level of He ($13.6<{h}\nu <24.59$).

Each run in the \code{serra} suite is initialized at $z=100$ from cosmological conditions generated with \code{music} \citep{hahn:2011mnras}, and zooms in on target DM halos selected at around $z\simeq 6$. The cosmic volume considered is $(20,{\rm Mpc}/{\rm h})^{3}$, evolved with a base grid of 8 levels. The zoom-in region has a volume of about $(2.1\,{\rm Mpc}/{\rm h})^{3}$ (approximately 5 times the virial radius of the target DM halo), and has 3 additional levels, resulting in a gas mass resolution of $m_b = 1.2\times 10^4 \msun$. Additionally, 6 extra levels of refinement are enabled in the zoom-in region based on a Lagrangian-like criterion, allowing the simulation to reach scales of $l_{\rm res}\simeq 30\,{\rm pc}$ at $z=6$ in the densest regions, similar to Galactic molecular clouds \citep[][]{Federrath+13}.


\subsection{Hyperspectral data cubes for [CII] and H$\alpha$ emission lines}

In \code{SERRA}, gas kinematics analyses involve two crucial post-processing steps: 1) line emission modelling \citep{vallini:2017, pallottini:2019} and 2) the generation of Hyperspectral Data Cubes \citep[HDCS,][]{Kohandel+19, Kohandel+20}. 

Due to the coarse nature of the chemical network used in hydrodynamical simulations, precise emission computation requires post-processing of the data to extract kinematic information. The line luminosity ($L^{\mathrm{em-line}}$) for each gas cell are obtained using the spectral synthesis code \code{CLOUDY} \citep{Ferland+17}. This process takes into account the interstellar radiation field, the turbulent and clumpy structure of the interstellar medium (ISM), which is parameterized as a function of the local gas Mach number \citep{vallini:2018, Pallottini+22}.

With information on $L^{\mathrm{em-line}}$, position ($\textbf{x}$), velocity ($\textbf{v}$), and thermal+turbulent line broadening ($(\sigmath^2 + \sigmant^2)^{1/2}$) for each gas cell within a specified field of view (FOV)\footnote{In this study, the FOV for the HDCs has a side of $2.5\,\rm{kpc}$.} and along a line of sight direction, we construct 3-dimensional HDCs.
These cubes comprise two spatial dimensions and one spectral dimension, effectively mapping the 6-dimensional data to coordinates ($x$, $y$, $v^{z}$). In HDCs, the surface brightness of the emission line is recorded for each voxel, providing valuable insights into the spatial and spectral distribution of the emission. The contribution of all gas cells within the FOV can be directly summed for optically thin emission lines. However, for optically thick lines, radiative transfer through dust needs to be considered when comparing with pre-dereddened observations \citep{behrens:2018dust}. 

In this study, we model the \CII~158$\mu$m emission line as a tracer of cold neutral/molecular gas and the nebular H$\alpha$ emission line as a tracer of warm ionized gas\footnote{While H$\alpha$ emission is optically thick, here we produce the non-attenuated HDCs and leave radiative transfer effects for future work.}.
Different gas phase tracers can yield different values of $V_{\mathrm{rot}}$ and $\sigma$ \citep{Kohandel+20, Ejdetjarn+22}. In this work, we estimate the velocity dispersion of a galaxy separately using two tracers: \CII~observations ($\sigmacii$) and H$\alpha$ ($\sigmahalpha$). Following a similar approach to \citet{Kohandel+20}, $\sigmaline$ represents the luminosity-weighted average velocity dispersion, calculated using the moment-2 and moment-0 maps of the corresponding emission line.
For the rotational velocity, we estimate\footnote{It is worth noting that this approximation is valid for a thin rotating disk with low velocity dispersion.} $V_{\mathrm{rot}}$ with the circular velocities of the galaxy, denoted as $v_{\rm c} = {{(G M_{\rm{dyn}}}/{r_d})}^{1/2}$, where $M_{\rm dyn}=M_{\rm g}+M_\star$ is the dynamical mass within the desired FOV, and $r_d$ is the disk effective radius, i.e. where $50\,\%$ of gas mass is contained.
In other words, we keep $V_{\mathrm{rot}}$ constant regardless of the tracer we use. Consequently, any differences observed in $V_{\mathrm{rot}}/\sigma$ for \CII~and H$\alpha$ synthetic observations arise due to disparities in $\sigmacii$ and $\sigmahalpha$. 

In this study, we focus on \code{SERRA} galaxies that exhibit a stellar mass of $10^{8}M_\odot$ and higher and covers the redshift range $4\leq z\leq 9$. Our sample includes 3218 galaxies with SFRs ranging from $0.04$ to $128$ $\mathrm{M_\odot \mathrm{yr}^{-1}}$ and stellar masses in $8 \le \log{M_\star/M_\odot} \le 10.3 $. We refer to Tab. \ref{tab:categories} for the overview of our sample. 

\begin{table}
\caption{Relevant properties of the simulated sample. Shown are the number of galaxies in each mass bin,  their average $V_{\rm{rot}}/\sigmacii$ and $V_{\rm{rot}}/\sigmahalpha$ ratios.\label{tab:categories}}
\begin{center}
\begin{tabular}{|l|c|c|c|}
\hline
$\rm{log}(M_\star/M_\odot)$  &  $\#$ of galaxies &$V_{\rm{rot}}/\sigma_{\mathrm{[CII]}}$&$V_{\rm{rot}}/\sigma_{\rm H\alpha}$\\
\hline
$\ge 10$& 142 & $8.5\pm 2.2$& $5.5\pm 3.2$ \\
$9-10$& 1149 & $5.0\pm 1.5$& $3.1\pm 1.3$ \\
$8-9$& 1927 & $4.1\pm 1.2$& $3.2\pm 1.1$ \\
\hline
\end{tabular}
\end{center}
\end{table}

\section{Rotation support in early galaxies}\label{sec:v-sigma-evol}

\begin{figure}
 \includegraphics[width=0.48\textwidth]{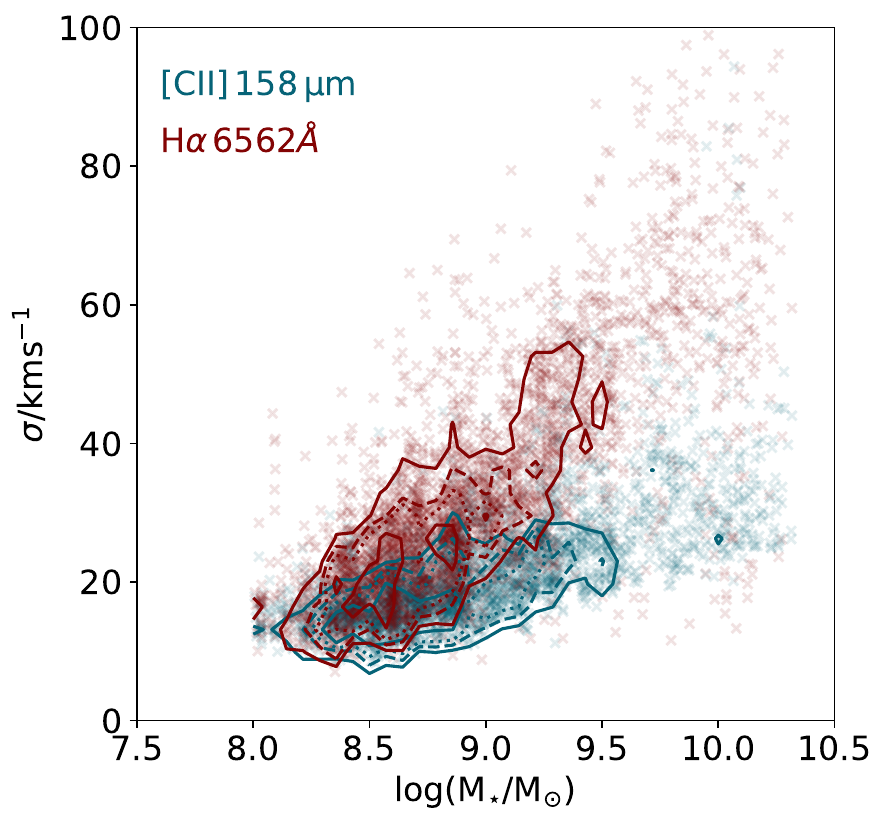}
\caption{The relationship between velocity dispersion ($\sigma$) and stellar mass ($M_\star$) in the \code{serra} galaxy sample: Blue (red) contours show the 1, 2, and 3-sigma probability density function levels for the $M_\star-\sigma$ relationship derived from synthetic \CII~ (H$\alpha$) observations. Individual data points are represented by crosses.
\label{fig:vsigma-mstar}}
\end{figure} 

\begin{figure*}
\includegraphics[width=0.9\textwidth]{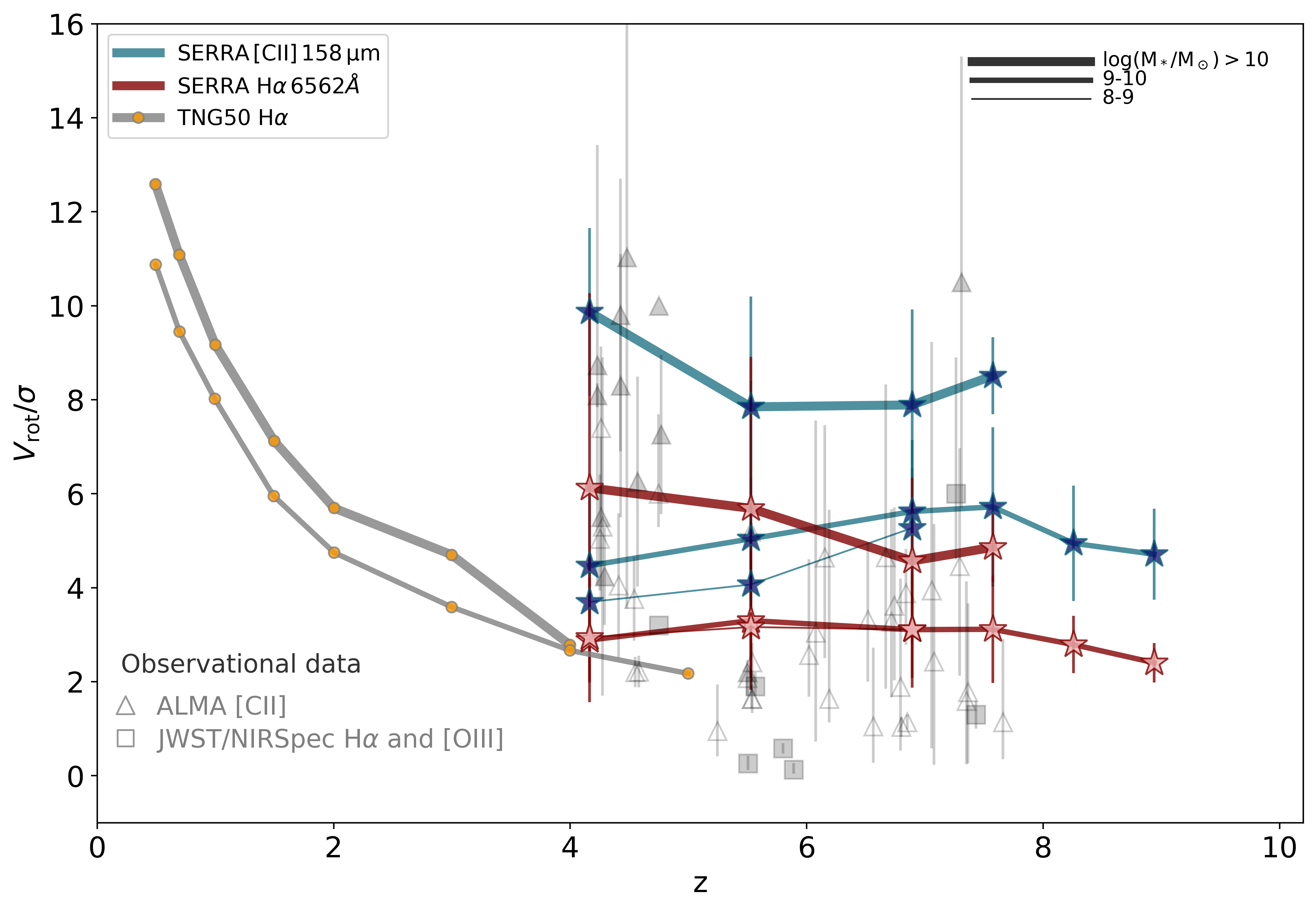}
\caption[v-sigma evolution]{
Redshift evolution of $V_{\rm{rot}}/\sigma$ in \code{SERRA}.
The three stellar mass bins are distinguished by line width; blue (red) markers represent $\sigma$ from \CII~(H$\alpha$) synthetic observations.
For comparison, predictions from \code{tng50} simulations \citep{Pillepich+19} are shown with orange-filled markers for $z<5$.
Grey triangles denote observational data points with \CII~line kinematics, including high-$z$ DCDs (\citealt{Rizzo+21,Lelli+21, Fraternali+21, Roman-Oliveira+23, Pope+23,Rowland+23}), ALPINE disk galaxies \citep{Jones+21}, archival $4<z<7.7$ \CII~data \citep{Parlanti+23}, the first discovered EoR disk galaxies \citep{Smit+18}, and the turbulent HZ4 system at $z=5.5$ \citep{Herrera-Camus+22}. Grey squares represent data with nebular H$\alpha$ and [OIII] kinematics \citep{DeGraaff+23, Parlanti+23b}. Filled (empty) markers distinguish spatially (barely) resolved observations. 
\label{fig:v-sigma-evol}
}
\end{figure*}

In Fig. \ref{fig:vsigma-mstar} and Fig. \ref{fig:v-sigma-evol}, we present the kinematic characteristics of our sample within the $M_\star$-$\sigma$ and $V_{\rm{rot}}/\sigma$-$z$\footnote{The redshift evolution for each mass category is obtained by dividing the category into 6 redshift bins between 4-9, and the markers indicate the mean of the sample in that redshift bin.} planes, respectively. In both cases, $\sigma$ values are derived from \CII~$158\,\mu\rm{m}$ and H$\alpha$ synthetic observations. In Fig. \ref{fig:v-sigma-evol}, along with \code{serra} galaxies, we plot predictions from \code{tng50} simulations \citep{Pillepich+19}\footnote{We have combined their original four mass bins of Fig. 14 in \citealt{Pillepich+19} into two composite bins: $\rm{log}(M_\star/M_\odot) > 10$ and $\rm{log}(M_\star/M_\odot) = 9-10$.}, as well as observed data of $z>4$ galaxies through FIR \CII~emission line by ALMA and nebular H$\alpha$ and \OIII~ lines by JWST/NIRspec. 

Regarding the $M_\star$-$\sigma$ relation, we find two distinct behaviours for the two tracers. Specifically, we find that $\sigmahalpha$ exhibits a steeper increase with $M_\star$ compared to $\sigmacii$. Interestingly, this trend is particularly pronounced in $\sim40\%$ of our high-mass bin galaxies and $30\%$  of those in the intermediate-mass bin, where $\sigmahalpha>2\sigmacii$. Such behaviour is not entirely unexpected, as even surveys of local galaxies (\citealt{Levy+18,Girard+21}) have highlighted systematic differences between hot ionized and neutral/molecular cold gas velocity dispersions. This disparity leads to a consistent difference between the results obtained using different tracers in \code{serra}, indicating that cold gas exhibits a higher degree of rotational support than warm ionized gas (see Tab. \ref{tab:categories} and Fig. \ref{fig:v-sigma-evol}). We will explore this point in more detail in Sec. \ref{sec:discussion}.

As for the $V_{\rm{rot}}/\sigma$-$z$, regardless of the kinematic tracer employed, we do not find any notable correlation even when different mass bins are considered. Nevertheless, a clear trend appears with the stellar mass of galaxies, indicating that most massive galaxies exhibit greater rotation support than the less massive ones. A milder version of such a correlation was also predicted in \code{tng50} for $z<4$ galaxies. 

\begin{table}
\caption{Dynamical categorization of \code{serra} galaxies.  Shown are the number of simulated galaxies in each dynamical stage depending on the adopted tracer.\label{tab:dynamical_stage}}
\begin{center}
\begin{tabular}{|l|c|c|c|}
\hline
Dynamical stage  &   $V_{\rm{rot}}/\sigma$ value&\CII & H$\alpha$\\
\hline
Super Cold & $\ge 10$& 37 & 25 \\ 
Cold &$4-10$ & 1926  & 540 \\
Warm &$2-4$ & 1242   & 2374\\
Hot &$\le 2$ & 13    & 279\\
\hline
\end{tabular}
\end{center}
\end{table}

The most noteworthy finding of our analysis is that when we classify our galaxies based on their $V_{\rm{rot}}/\sigma$ ratio (as outlined in Tab. \ref{tab:dynamical_stage}), not only we identify Super Cold disks ($V_{\rm{rot}}/\sigma>10$ similar to SFGs observed at $z\sim 4$ \citep{Rizzo+21, Fraternali+21} and at $z\sim 7$ at \citealt{Rowland+23}) within our massive sub-sample ($M_\star> 10^{10}\,M_\odot$), but we also ascertain that \CII~emitting gas in $\sim 60\%$ of the whole sample is dynamically cold, having $4<V_{\rm{rot}}/\sigma<10$.
This finding suggests that galactic disks can form as early as the EoR and if deep ALMA observations targeting SFGs at $z>4$ become available, more dynamically cold disks will likely be uncovered. 

For the majority of galaxies with \CII~observations, $V_{\rm{rot}}/\sigma$ values at $z>5$ fall below the predicted mean values from \code{serra} simulations. This discrepancy can primarily be attributed to the marginal resolution of these observations ($\sim 0.1-1.5$ arcsecs), compared to our synthetic datacubes featuring a higher spatial resolution of $0.005$ arcsecs. In fact, in \citealt{Kohandel+20}, we showed how the beam-smearing effect in low-resolution observations can lead to a substantial overestimate of velocity dispersion, reaching up to $\sim 100\%$ (see also \citealt{Ejdetjarn+22, Rizzo+22}). Another potential cause of such discrepancy is the challenge of accurately estimating disk inclination in high-$z$ kinematic observations. This becomes crucial when determining kinematic properties, such as velocity dispersion, from integrated spectra, where the shape heavily correlates with the disk inclination \citep{Kohandel+19}.

\section{Discussion}\label{sec:discussion}

\begin{figure*}
\includegraphics[width=\textwidth]%
{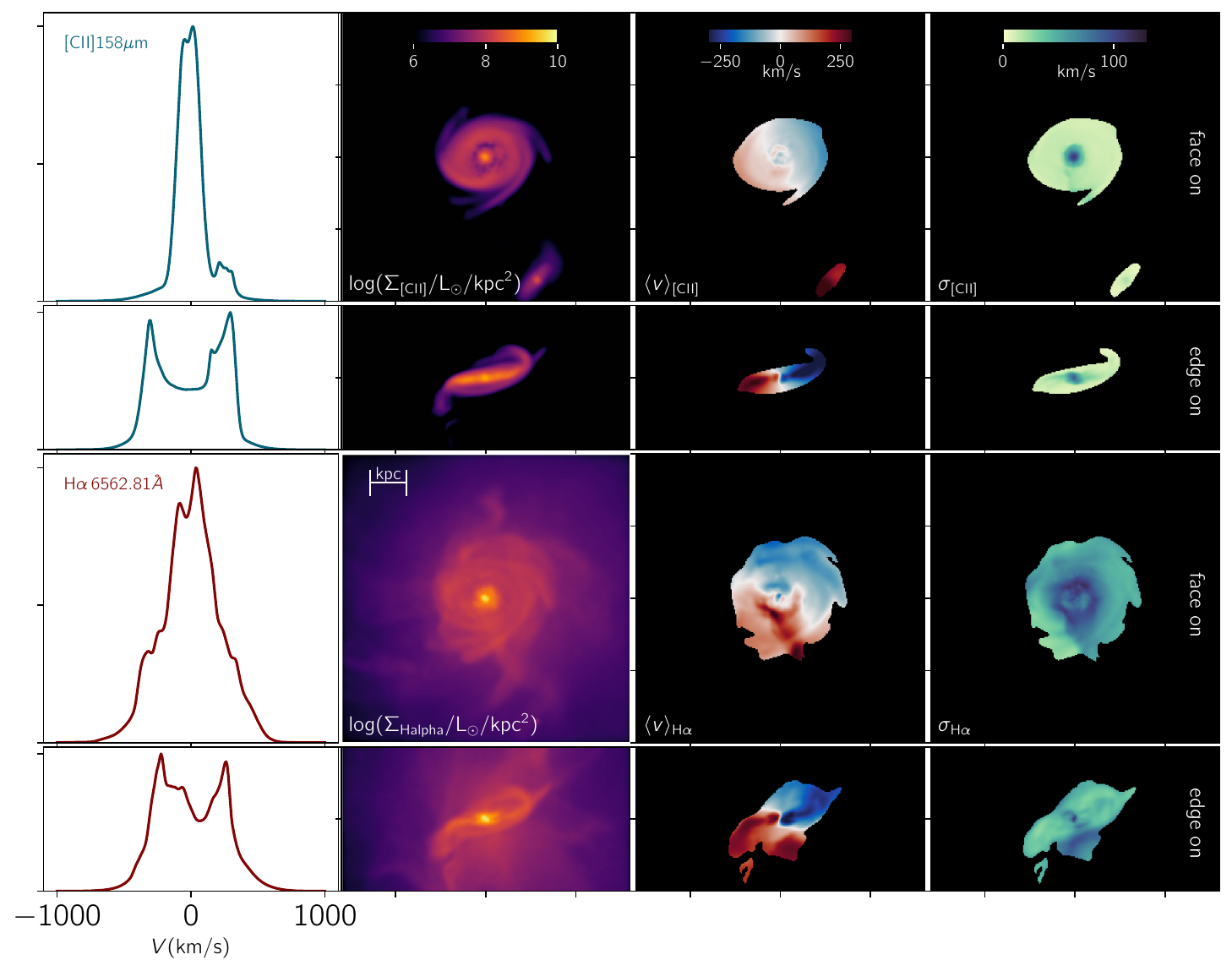}
\caption[Hibiscus]{Multi-wavelength kinematics of \quotes{Hibiscus} at $z=4.5$. In the top (last) two rows we display FIR \CII~158$\mu$m (nebular H$\alpha$) syntactic observation for line spectra, 0, 1, and 2 moment maps in different columns, for both face-on and edge-on views.
\label{fig:hibiscus}
}
\end{figure*}

To better clarify the above findings, we now focus on two representative galaxies in our massive sub-sample: \quotes{Hibiscus} with $M_\star = 1.5\times 10^{10}\,M_\odot$ and $\rm{SFR}=16M_\odot\rm{yr}^{-1}$ at $z = 4.5$ and \quotes{Narcissus} with $M_\star = 1.2\times 10^{10}\,M_{\odot}$ and a $\rm{SFR}= 52\,M_\odot \rm{yr}^{-1}$ at $z=6.8$. These galaxies have similar \CII~kinematics (i.e., $V_{\rm{rot}}/\sigmacii>10$), but their H$\alpha$ kinematics is very different.

\subsection{Why [CII] and H$\alpha$ kinematics are different?}

In Fig. \ref{fig:hibiscus}, we show the synthetic \CII~and H$\alpha$ maps and kinematics observables for face-on and edge-on views of Hibiscus. 
The \CII~spectrum is characterized by a narrow and prominently Gaussian-shaped profile (FWHM$\sim 185\,\kms$), while the H$\alpha$ spectrum appears more complex, broader (FWHM$\sim 437\,\kms$), and exhibits high-velocity wings. Such broad wings in the spectrum could be indicative of outflowing gas.
Indeed, comparing the moment maps, we see that the \CII~emission line effectively traces Hibiscus thin gaseous disk, while H$\alpha$ traces ionized gas that lies beyond the disk plane, including gas that might be in an in/outflowing state.
Such a difference between various phases of the ISM could conceivably arise due to distinct effects of stellar feedback influencing them, as suggested by isolated disk galaxy simulations \citep{Ejdetjarn+22}.
This illustrates that the observed velocity dispersion in a given galaxy using H$\alpha$ data may not solely arise from turbulence within the galactic disks. Instead, a substantial contribution from outflows may be in effect, introducing an additional layer of complexity to the data interpretation.

We note that the spatial extent of \CII~and H$\alpha$ emission in Hibiscus differs significantly. The \CII~ is 4$\times$ more extended than the stellar effective radius, similar to observed high-$z$ galaxies \citep{Fujimoto+19,Carniani+20,Fudamoto+22}. The H$\alpha$ distribution is even more far-flung. This happens since H$\alpha$ originates from the $T \sim 10^4\,$K photoionized regions outside the disk that are part of an expanding, cooling outflow through which LyC photons percolate. As carbon in these regions is ionized to higher states (e.g. CIII), \CII~emission is limited to denser, more confined regions where recombination rates are higher. Interestingly, this shows that H$\alpha$ halos are intimately linked to the presence and morphology of these outflows, offering intriguing prospects for their detection with JWST. 

Lastly, it is important to highlight that \CII~kinematics can also be challenging and merits deeper exploration. In particular, observations of the so-called \CII~halos \citep{gallerani:2018outflow, Fujimoto+19, Fujimoto+21, Ginolfi+20} have been interpreted in the framework of outflow models \citep{Pizzati+20, Pizzati+23}, but they have not yet been reproduced by cosmological simulations \citep{Fujimoto+19, Arata+20}. 

\subsection{Are high-$z$ dynamically cold disks a transient feature?}

\begin{figure}
\includegraphics[width=0.5\textwidth]{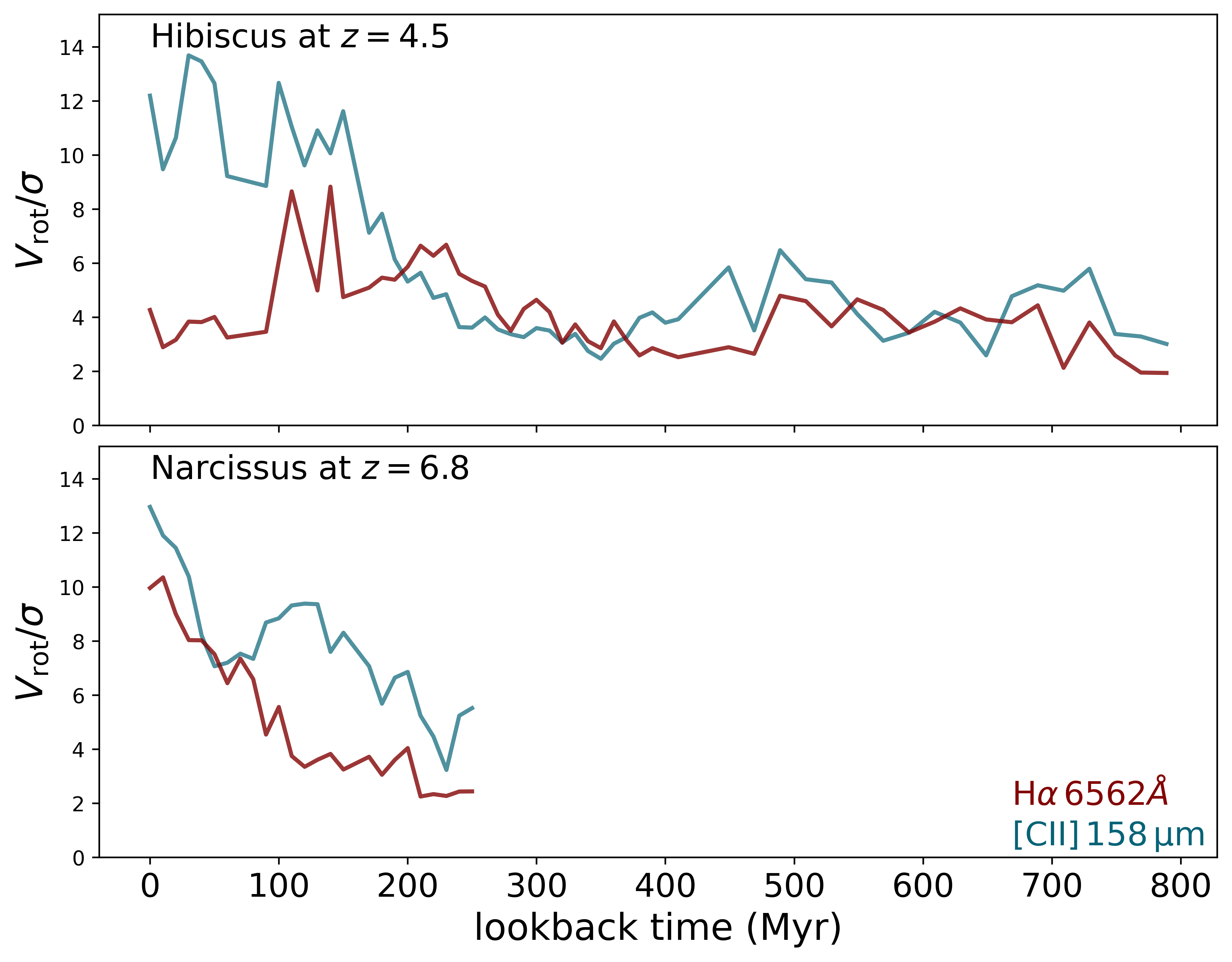}
\caption[Narcicus]{Example of dynamical evolution of two \code{SERRA} galaxies traced by [CII] and H$\alpha$; at $z=4.5$, Hibiscus (top) has a [CII] dynamically cold disk and a turbulent H$\alpha$ emitting gas possibly featuring galactic outflows (see also Fig. \ref{fig:hibiscus}); at $z=6.8$, Narcisuss (bottom) appears as a super cold galaxy both in \CII~and H$\alpha$. \label{fig:evolution-hibiscus-narcissus}}
\end{figure}

To explore the stability of cold disks, we take a closer look at the evolutionary path undertaken by individual galaxies.
In Fig. \ref{fig:evolution-hibiscus-narcissus}, we show the evolution of $V_{\rm{rot}}/\sigma$ as a function of lookback time for Hibiscus and Narcissus. We see that, considering only \CII~emitting gas, both galaxies exhibit a consistent $V_{\rm{rot}}/\sigma$ around $2-4$ up until roughly $200\,\rm{Myr}$. During this interval, the rotation support rises due to the effective accretion of gas and the efficient transfer of angular momentum into the disk. Indeed, if we estimate the disk orbital time $t_{\rm{orb}} = 2\pi r_{d}/V_{\rm{rot}}$, for Hibiscus it is $\sim 16\,\rm{Myr}$, and $\sim 21\,\rm{Myr}$ for Narcissus. Therefore, these high-$z$ DCDs survive for more than 10 orbital times.

Regarding the $V_{\rm{rot}}/\sigmahalpha$, there is an interesting difference between the two systems. As witnessed in Fig. \ref{fig:hibiscus}, the H$\alpha$ emitting gas is found to be dynamically warm in the last $\sim 200\, \rm Myr$ for Hibiscus, potentially attributable to the presence of outflows. 
%
However, in the case of Narcissus, the gas traced by both emission lines follows a similar evolutionary path. Despite a slightly lower $\sigmacii$ than $\sigmahalpha$, this galaxy remains dynamically cold according to both tracers.
Considering the comparable stellar masses of these galaxies, differences in their star formation histories, feedback effects, or other global properties may be responsible for the different behaviour of the tracers.
%

\section{Conclusions}

In this {\it Letter}, we investigate the possible existence of dynamically cold disks (with significant rotation support) in the early Universe using a sample of 3218 Lyman Break Galaxies (LBG) from the \code{SERRA} zoom-in cosmological simulations.
We analyze the kinematic of both \CII~and H$\alpha$ in the redshift range of $4\le z \le 9$ for LBGs with $8\le\log{(M_\star/M_\odot)}\le 10.3$ and $0< \rm{SFR} \le 128$. Our main conclusions are the following.

\begin{itemize}
    \item[$\bullet$] A strong (mild) correlation is present between stellar mass and gas velocity dispersion when H$\alpha$ (\CII) synthetic observations are considered. The difference arises mostly for $M_\star > 10^{9}M_\odot$ galaxies where  $\sigmahalpha>2\sigmacii$.  
    \item[$\bullet$] Irrespective of galaxy mass and the chosen kinematic tracer, our analysis reveals no significant redshift dependence in the ratio $V_{\rm{rot}}/\sigma$.
    \item[$\bullet$] Massive ($M_\star \ge 10^{10}\,M_\odot$) galaxies in \code{serra} settle into dynamically Super Cold disks with $V_{\rm{rot}}/\sigma>10$ at $z>4$. Such cold disks are not transient features as they last for more than 10 galaxy orbital times ($\sim 200\,\rm{Myr}$).
\end{itemize}

We have shown that in \code{serra} galaxies, \CII~ effectively traces the thin gaseous disks within galaxies, while H$\alpha$ emission can also trace the ionized gas outside the disk. The differences in the kinematics of \CII~and H$\alpha$ may be attributed to galactic outflows, although further exploration is necessary to substantiate and statistically quantify this point. We show that the identification of H$\alpha$ halos could be a signature of such galactic outflows.
We foresee that more high-$z$ dynamically cold disks will be found with the increasing availability of deep ALMA observations targeting \CII~$158\,\mu\rm{m}$ in galaxies with stellar masses exceeding $10^{9}M_\odot$.

In view of the essential role of multiple tracers in gaining a comprehensive understanding of early galaxy kinematics, we emphasize that the ALMA-JWST/NIRspec synergy will be essential.

\begin{acknowledgements}
MK and AF acknowledge support from the ERC Advanced Grant INTERSTELLAR H2020/740120 (PI: Ferrara).
Any dissemination of results must indicate that it reflects only the author's view and that the Commission is not responsible for any use that may be made of the information it contains.
We acknowledge the CINECA award under the ISCRA initiative for the availability of high-performance computing resources and support from the Class B project SERRA HP10BPUZ8F (PI: Pallottini).
We gratefully acknowledge the computational resources of the Center for High-Performance Computing (CHPC) at SNS.
We acknowledge use of the Python programming language \citep{VanRossum1991}, Astropy \citep{astropy}, Cython \citep{behnel2010cython}, Matplotlib \citep{Hunter2007}, NumPy \citep{VanDerWalt2011}, \code{pynbody} \citep{pynbody}, and SciPy \citep{scipy2019}.

\end{acknowledgements}

%
%
\bibliographystyle{aa_url}
\bibliography{bibliography/master,bibliography/codes} 

\end{document}